\documentclass[aps,prl,twocolumn,showpacs,superscriptaddress,showkeys]{revtex4-1}
\usepackage{amssymb}
\usepackage{amsmath}
\usepackage{amsfonts}
\usepackage{graphicx}
\usepackage{xcolor}
\usepackage{xspace}
\usepackage{ulem}
\usepackage{mathtools}
\usepackage{hhline}

\usepackage[T1]{fontenc}
\usepackage{newtxtext}
\usepackage[varg]{newtxmath}

\newcommand{\AddrAve}{Departamento de F\'{\i}sica da Universidade de Aveiro and CIDMA,  Campus de Santiago, 3810-183 Aveiro, Portugal}

\begin{document}


\title{Warm Little Inflaton becomes Cold Dark Matter}

\author{Jo\~{a}o G.~Rosa} \email{joao.rosa@ua.pt} \affiliation{\AddrAve}
\author{Lu\'{\i}s B. Ventura} \email{lbventura@ua.pt} \affiliation{\AddrAve}

\date{\today}

\begin{abstract}
We present a model where the inflaton can naturally account for all the dark matter in the Universe within the warm inflation paradigm. In particular, we show that the symmetries and particle content of the Warm Little Inflaton scenario (i) avoid large thermal and radiative corrections to the scalar potential, (ii) allow for sufficiently strong dissipative effects to sustain a radiation bath during inflation that becomes dominant at the end of the slow-roll regime, and (iii) enable a stable inflaton remnant in the post-inflationary epochs. The latter behaves as dark radiation during nucleosynthesis, leading to a non-negligible contribution to the effective number of relativistic degrees of freedom, and becomes the dominant cold dark matter component in the Universe shortly before matter-radiation equality for inflaton masses in the $10^{-4}-10^{-1}$ eV range. Cold dark matter isocurvature perturbations, anti-correlated with the main adiabatic component, provide a smoking gun for this scenario that can be tested in the near future.
\end{abstract}

.


\maketitle


Inflation \cite{inflation} has inevitably become a part of the modern cosmological paradigm. Observations are consistent with its predictions of a flat universe with a nearly scale-invariant, Gaussian and adiabatic spectrum of primordial density perturbations \cite{PlanckCI:2018}. However, the exact nature of the inflaton field and its scalar potential remain open questions, that may hopefully be addressed with future measurements of B-modes in the cosmic microwave background (CMB) polarization, as well as possibly non-Gaussian features and isocurvature perturbations.

The present ambiguity is no less apparent when one regards two consistent, yet distinct, descriptions of this period of accelerated expansion: cold and warm inflation \cite{Berera:1995wh, Berera:1995ie}. The former is described by the overdamped motion of one (or more) field(s), coupled solely with gravity, whose nearly flat potential dominates the energy content of the Universe, acting as an effective cosmological constant for a finite period. Inflaton quantum fluctuations then act as seeds for structure formation and CMB anisotropies.

If, however, interactions between the inflaton, $\phi$, and other fields are non-negligible during inflation, dissipative effects transfer the inflaton's energy into light degrees of freedom (DOF), sustaining a sub-dominant radiation bath that significantly changes the dynamics of inflation \cite{Berera:2008ar}. In particular, the additional dissipation makes the inflaton field roll more slowly than in the cold case \cite{Swampland}. When dissipation becomes sufficiently strong, the radiation bath can smoothly take over as the dominant component at the end of inflation, with no need for a reheating period \cite{BasteroGil:2009ec, Berera:1996fm}. Since dissipation sources small thermal fluctuations of the inflaton field about the background value $\phi$, the primordial spectrum of curvature perturbations can differ significantly in the cold and warm inflation scenarios \cite{Berera:1999ws, Hall:2003zp, Moss:2007cv, Graham:2009bf, Ramos:2013nsa, Bartrum:2013oka, Bartrum:2013fia, Benetti:2016jhf}, the latter e.g. predicting a suppression of the tensor-to-scalar ratio in chaotic models \cite{BasteroGil:2009ec,Berera:1999ws,Bartrum:2013fia}. Warm inflation thus provides a unique observational window into inflationary particle physics.

Warm inflation models were, however, hindered by several technical difficulties \cite{BGR, YL}. Interactions between the inflaton and other fields, e.g.  $g^2 \phi^2 \chi^2$ or $g \phi \overline{\psi}\psi$, typically give the latter a large mass. On the one hand, to keep the fields light, and also avoid the associated large thermal corrections to the inflaton potential, one must consider small coupling constants that make dissipative effects too feeble to sustain a radiation bath for $\sim$50-60 e-folds of inflation. On the other hand, heavy fields acting as mediators between the inflaton and light degrees of freedom in the radiation bath can yield sufficiently strong dissipative effects \cite{Berera:2002sp, Moss:2006gt, BasteroGil:2010pb, BasteroGil:2012cm}, but at the expense of considering a large number of such mediators that may only be present in certain string constructions \cite{BasteroGil:2011mr} or extra-dimensional models \cite{Matsuda:2012kc}.

These shortcomings were recently overcome in the \textit{Warm Little Inflaton} (WLI) scenario, which provides the only consistent model of warm inflation with only two light fields coupled to the inflaton \cite{Bastero-Gil:2016}. In this Letter, we show, for the first time, that the underlying symmetries and particle content of this model necessarily lead to a stable inflaton remnant that survives until the present day, naturally accounting for the cold dark matter component in our Universe. In this sense, warm inflation implies inflaton-dark matter unification. 

This is hard to implement within the cold inflation paradigm because the inflaton must decay efficiently at the end of inflation to ensure a successful ``reheating" of the Universe, but cannot do so completely to provide a sufficiently long-lived dark relic \cite{Liddle:2008}. This can be achieved by introducing additional symmetries or cosmological phase(s) which, however, do not affect the inflationary dynamics and have, in general, no direct observational imprint \cite{thermal,th_infl,kinetic,regeneration,singlet,Daido:2017wwb, Hooper:2018buz,Bastero-Gil:2015lga}.

In this Letter, we propose a radically different unification scenario, where the {\it inflaton decays during, but not after, inflation}. We show that this is possible due to the very same symmetries that protect the scalar potential against thermal and quantum corrections, independently of its form, while allowing for sufficiently strong adiabatic dissipation effects that become exponentially suppressed once the inflaton exits the slow-roll regime and radiation smoothly takes over. These symmetries then ensure that the inflaton behaves as a stable cold relic while oscillating about the minimum of its potential at late times, regardless of the full form of the scalar potential. 

The WLI model includes two complex scalar fields, $\phi_{1,2}$, with equal charge $q$ under a U(1) gauge symmetry that is spontaneously broken by their identical vacuum expectation values, $\langle \phi_1\rangle=\langle \phi_2\rangle\equiv M/\sqrt{2}$. This yields masses of order $M$ for the radial scalar components and the U(1) gauge field that then decouple from the dynamics for temperatures $T\lesssim M$. The remaining physical light degree of freedom is the relative phase between $\phi_1$ and $\phi_2$, which we identify as the inflaton field, $\phi$:
\begin{equation}
	\phi_1 = \frac{M}{\sqrt{2}}e^{i \phi/M} \quad, \quad \phi_2 = \frac{M}{\sqrt{2}}e^{-i \phi/M} \quad.
\end{equation}
 The model also includes a pair of fermions, $\psi_1$ and $\psi_2$, whose left-handed (right-handed) components have U(1) charge $q$ (0), and we impose an interchange symmetry $\phi_1\leftrightarrow \phi_2$, $\psi_1\leftrightarrow \psi_2$, such that the allowed Yukawa interactions are given by:
\begin{eqnarray}
-\mathcal{L}_{\phi \psi} &=& \frac{1}{\sqrt{2}}g \phi_1 \overline{\psi}_{1L}\psi_{1R} 
+ \frac{1}{\sqrt{2}}g \phi_2 \overline{\psi}_{2L}\psi_{2R} + \text{H.c.} =
\nonumber\\
&=& gM \overline{\psi}_1e^{i \gamma_5 \phi/M}\psi_1 + gM \overline{\psi}_2 e^{-i \gamma_5 \phi/M}\psi_2 \: \:, \label{mWLI Lagrangian}
\end{eqnarray}
where we have also imposed the sequestering of the ``1" and ``2" sectors, which may be achieved e.g.~by considering additional  global symmetries for each sector or by physically separating them along an extra compact dimension. This is in contrast with the original WLI proposal \cite{Bastero-Gil:2016}, providing the additional appealing feature that the fermion masses, $m_1 = m_2 = gM$, and therefore the associated radiative and thermal corrections to the effective potential are independent of the inflaton field. Note that if $\phi$ were constant throughout the whole space-time, one could remove it completely from the fermion Lagrangian through a chiral rotation. Hence, the only effects of the inflaton field's interactions with the fermions are related to its dynamical nature, i.e.~they correspond to non-local contributions to the effective action, in particular dissipative effects.

The interchange symmetry corresponds to a $\mathbb{Z}_2$ reflection for the inflaton field, $\phi\leftrightarrow -\phi$, that protects it from decaying into any other fields besides the fermions $\psi_{1,2}$. As discussed below, a significant dissipation of the inflaton's energy implies $gM \lesssim T \lesssim M$, which is only satisfied {\it during} inflation, thus ensuring the stability of the inflaton in the post-inflationary Universe. 

The fermion fields may decay into light scalar and fermion fields, $\sigma$ and $\psi_\sigma$, respectively, with appropriate U(1) charges, through Yukawa interactions:
\begin{equation}
-\mathcal{L}_{\psi \sigma} = -h \sigma \sum_{i=1,2}(\overline{\psi}_{iL}\psi_{\sigma R}+\overline{\psi}_{\sigma L}\psi_{iR}) \quad.
\end{equation}
The dissipation coefficient resulting from the inflaton-fermion interactions can be computed using standard thermal field theory tools in the adiabatic regime \cite{Berera:2008ar,  BasteroGil:2010pb}, where the fermions are kept close to thermal equilibrium through decays and inverse decays, $\Gamma_{\psi}\gtrsim H\gg |\dot{\phi}/\phi|$ \cite{note3a}. Its dominant contribution corresponds to on-shell fermion production \cite{BasteroGil:2012cm}, being given by:
\begin{equation}
\Upsilon = 4\frac{g^2}{T} \int \frac{\text{d}^3 p}{(2\pi)^3} \frac{\Gamma_{\psi}}{m_{\psi}^2} n_F(\omega_p) (1 - n_F(\omega_p)) \quad,
\end{equation}
where $m_{\psi} = \sqrt{g^2 M^2 + h^2 T^2 /8}$ is the thermally corrected fermion mass, $\omega_p = \sqrt{|\mathbf{p}|^2 + m_{\psi}^2}$, $\Gamma_{\psi}$ is the fermion decay width (see \cite{Bastero-Gil:2016}) and $n_F$ is the Fermi-Dirac distribution. The momentum integral can be computed analytically for $m_{\psi}/T \ll 1$ and $m_{\psi}/T \gg 1$, yielding:
\begin{equation}
\begin{aligned}
&\Upsilon \approx  \frac{2 g^2 h^2}{(2\pi)^3} T \left(\frac{1}{5} - \ln \left( \frac{m_{\psi}}{T}\right)\right)\: \:, \: \:  m_{\psi}/T  \ll 1 \quad,
\\
&\Upsilon \approx  \frac{2 g^2 h^2}{(2\pi)^3} T \sqrt{\frac{\pi}{2}}  \sqrt{\frac{m_{\psi}}{T}} e^{-\frac{m_{\psi}}{T}} \: \:, \: \:  m_{\psi}/T  \gg 1 \quad. \label{LowTDC}
\end{aligned}
\end{equation}
In the high temperature regime relevant during inflation, the dissipation coefficient is proportional to the temperature as in the original WLI model, although the proportionality constant is smaller in this case. Inflationary observables can thus be computed as in \cite{Bastero-Gil:2016, Bastero-Gil:2017wwl, Bastero-Gil:2018uep}, to which we refer the interested reader. After inflation, dissipative effects are exponentially suppressed as the $\psi_{1,2}$ fermions become non-relativistic, halting the energy transfer between the inflaton condensate $\phi$ and the light DOF.

The evolution of the background homogeneous inflaton field and entropy density is then dictated by:
\begin{align}
\ddot{\phi} + (3H +\Upsilon)\dot{\phi} + V_{,\phi} = 0~,\qquad
\dot{s} + 3H s = \frac{\Upsilon \dot{\phi}^2}{T}~,
\end{align}
alongside the Friedmann equation $H^2 = (\rho_{\phi} + \rho_r)/(3 M_{\text{P}}^2)$, where $\rho_{\phi} = (\dot{\phi}^2 /2) + V(\phi)$, $s = (2\pi^2 /45) g_* T^3= 4\rho_r/(3T)$ and $g_* $ is the number of relativistic DOF. Since the inflaton is a gauge singlet, the scalar potential is an arbitrary even function of the field, so we consider the simplest renormalizable potential, $V(\phi) =  m_{\phi}^2 \phi^2/2 + \lambda \phi^4$, with the quartic (quadratic) term dominating at large (small) field values. 
\begin{figure}[t]
	\centering\includegraphics[scale=0.42]{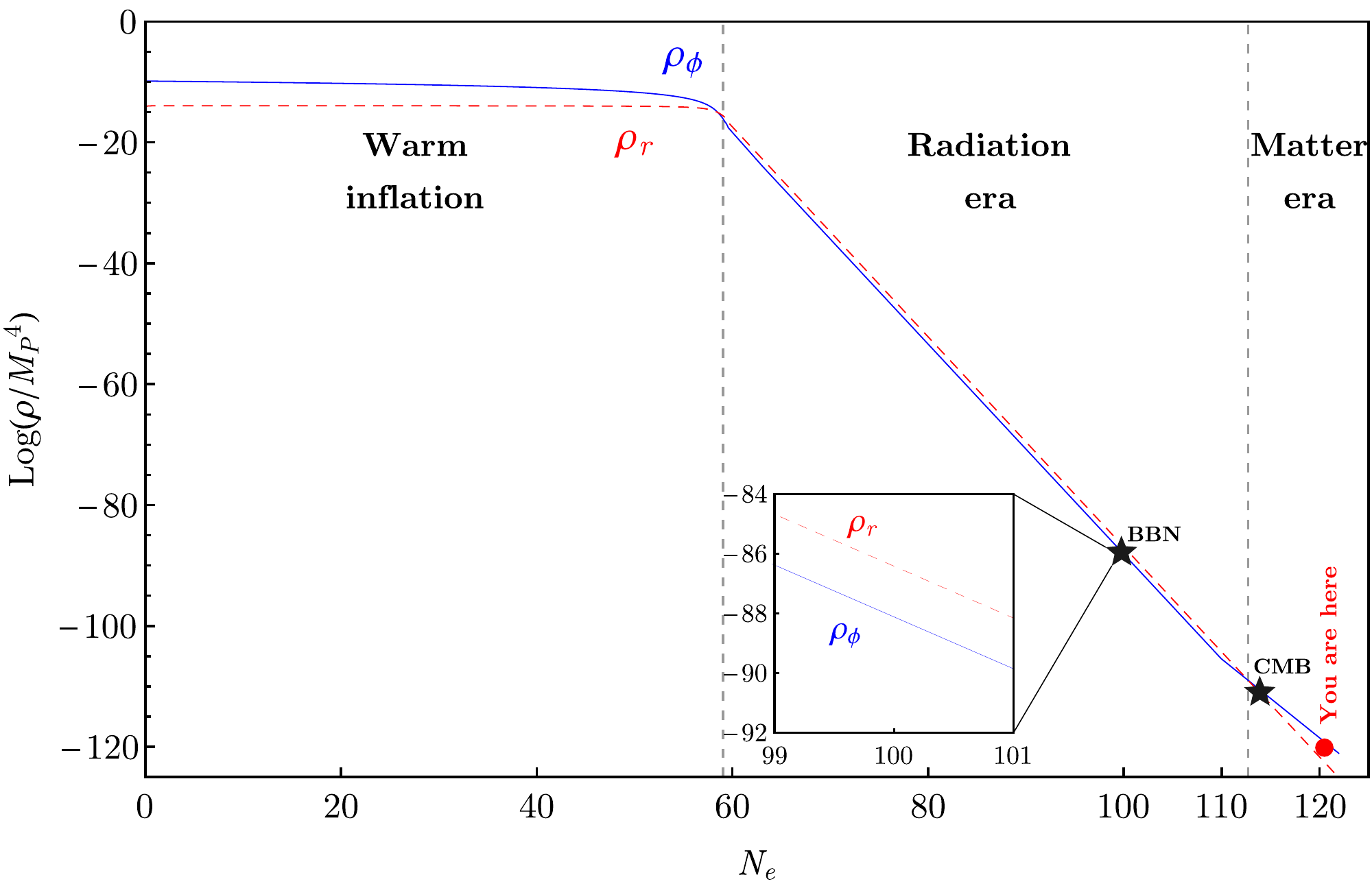}
\caption{Cosmological evolution of the inflaton (solid blue line) and radiation (dashed red line) energy densities for a representative choice of parameters. The vertical dashed black lines mark the inflaton-radiation equality times at the end of inflation and of the radiation era. The inset plot shows that the inflaton behaves as a sub-dominant dark radiation component during nucleosynthesis (BBN) ($T \sim 1 \text{ MeV}$)}. The last scattering surface (CMB) ($T \sim 0.3 \text{ eV}$) and the present day are also highlighted.
\label{cosmological_evolution}
\end{figure}

\begin{figure}
	\centering\includegraphics[scale=0.5]{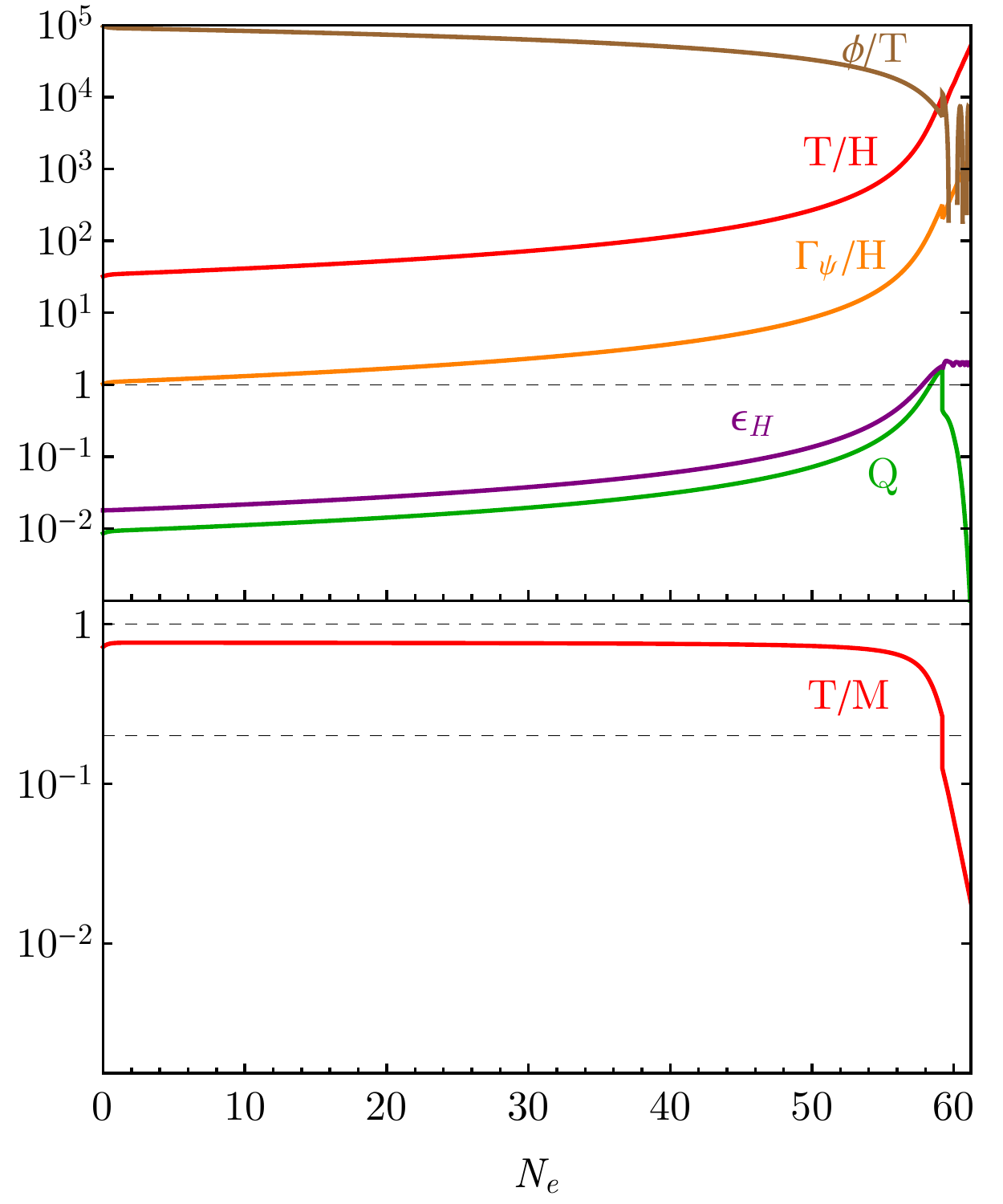}
	\caption{Evolution of $\phi/T$ (brown), $T/H$ (red), $\Gamma_{\psi}/H$ (orange), $Q$ (green), $\epsilon_{H} \equiv - \dot{H}/H^2$, $Q \equiv \Upsilon/3H$ (purple)} and $T/M$ (red, bottom plot) during inflation. In this example, $h=1.8$, $g=0.2$, $M= 7.2 \times 10^{14}$ GeV, with horizon-crossing conditions $\phi_* = 21 M_{\text{P}}$, $Q_* = 0.0072$. This yields $58$ e-folds of inflation, a scalar spectral index $n_s = 0.965$ and a tensor-to-scalar ratio $r = 4.3 \times 10^{-3}$ for nearly-thermal inflaton fluctuations (see \cite{Bastero-Gil:2016, Bastero-Gil:2017wwl, Bastero-Gil:2018uep}).
	\label{inflationary_dynamics}
\end{figure}

A representative example of the cosmological dynamics is illustrated in Figs.~\ref{cosmological_evolution} and \ref{inflationary_dynamics}. The inflaton potential energy dominates the energy balance for $\sim$60 e-folds in the slow-roll regime, while sustaining a nearly constant radiation abundance through dissipative effects. These become strong ($Q\gtrsim 1$) at the end of inflation, allowing for radiation to take over as the dominant component, at a temperature $T_R\sim 10^{13}-10^{14}$ GeV. Almost simultaneously, the temperature drops below the mass of the $\psi_{1,2}$ fermions and they gradually decouple from the dynamics, therefore making the inflaton effectively stable as we discuss below. At this stage, the ratio $T/H\sim 10^4$ becomes large enough to excite all the Standard Model DOF \cite{KolbTurner}, resulting in a temperature drop by a factor $\sim$2, further suppressing dissipative effects \cite{note4a}.

In the absence of significant dissipation, the inflaton condensate starts oscillating about the origin in the quartic potential, with amplitude $\phi\propto a^{-1}$, hence behaving as dark radiation. This phase lasts until the amplitude drops below $\phi_{\text{DM}} = m_{\phi}/\sqrt{2 \lambda}$, after which the quadratic term dominates and the inflaton remnant behaves as cold (pressureless) dark matter until the present day.

The ratio $\phi/T$ remains constant throughout the dark radiation phase (up to changes in $g_*$), and we find numerically that it is an $\mathcal{O}(1-10)$ factor below its value at inflaton-radiation equality, $\phi/T \sim 10^4$. This suppression is the result of the faster decay of the inflaton amplitude compared to the temperature at the onset of inflaton oscillations. This assumes that the inflaton field does not decay into or scatter significantly off the particles in the thermal bath after inflation. The latter could, in particular, thermalize the inflaton particles that would later decouple from the thermal bath as a standard WIMP. However, since the temperature falls below the mass of the fermions just after inflation, inflaton interactions with the remaining light fields are significantly suppressed.

Consider first the 4-body decay of inflaton, $\phi \rightarrow \overline{\psi}_{\sigma} \psi_{\sigma} \sigma \sigma $, mediated by the heavy $\psi_{1,2}$ fermions, for which we obtain, in the dark radiation phase where the effective inflaton mass $m_{\phi, \text{eff}}\approx \sqrt{12\lambda}\phi$ \cite{Asatrian:2013}:
\begin{eqnarray}
{\Gamma_{\phi}\over H}&\approx &10^{-19} {h^4\over g^2}\left({\lambda\over10^{-15}}\right)^{5/2} \left(\frac{\phi}{M_P}\right)^3\left(\frac{\phi/T}{10^4}\right)^2\nonumber\\
&\times&\left(\frac{10^{-4}M_P}{M}\right)^4 \sin^2\left({\phi\over M}\right)~,
\end{eqnarray}
which vanishes at the origin due to the underlying $\mathbb{Z}_2$ symmetry, and is in fact negligible throughout the whole radiation era in the parametric range relevant for a successful inflationary dynamics, as in the example of Fig.~\ref{inflationary_dynamics}. 

Low-momentum $\phi$-particles in the inflaton condensate ($|\mathbf{p}_{\phi}| \lesssim H$) can also scatter off light particles in the thermal bath, e.g.~$\phi \sigma \rightarrow \phi \sigma \overline{\psi}_{\sigma} \psi_{\sigma}$ and similar processes that could lead to the condensate's evaporation and thermalization of the inflaton particles. This process is again mediated by the heavy $\psi_{1,2}$ fermions, with center of mass energies $s \approx 2 m_{\phi,\text{eff}} T$, yielding:
\begin{eqnarray}
{\Gamma_{\phi\sigma}\over H}	& \approx& 10^{-9} {h^4 g}\! \left(\frac{\lambda}{10^{-15}}\right)
\left(\frac{\phi}{M_P}\right)^{2} \left(\frac{T}{gM}\right)^{3} \nonumber\\
&\times&\left(\frac{10^{-4}M_P}{M}\right)^{3}\cos^2\left({\phi\over M}\right)~, \label{Scat Ratio}
\end{eqnarray}
which is also negligible in the relevant parametric range.

Both the 4-body decay and scattering processes discussed previously are suppressed by the mass of the $\psi_{1,2}$ fermions, which is larger than the temperature in the radiation era ($T<T_R$). One must, however, note that direct couplings between the inflaton and the light $\sigma$ and $\psi_\sigma$ fields are not forbidden by any symmetries and are generated radiatively through loop diagrams involving $\psi_1$ and $\psi_2$. Decay processes such as $\phi\rightarrow \sigma\sigma,~ \overline{\psi}_\sigma \psi_\sigma$ may thus be allowed, although they become gradually forbidden as $\phi \rightarrow 0$ due to the $\mathbb{Z}_2$ symmetry. Similarly, these radiatively generated vertices may also lead to 2-body scatterings that could thermalize the inflaton condensate. 

One can nevertheless consider the particular case where the renormalized values of such effective vertices are small for $gM\lesssim T \lesssim M$ during inflation, such that they do not affect the inflationary dynamics. The Appelquist-Carazzone decoupling theorem \cite{Appelquist:1974tg} then ensures that they will not run significantly below the $\psi_{1,2}$ mass threshold, hence suppressing their effects in the post-inflationary epoch (see also e.g.~\cite{BasteroGil:2010vq}). We will consider this case in the remainder of our discussion, keeping in mind that, in more general scenarios, the inflaton may decay in the radiation era, although never completely due to the $\mathbb{Z}_2$ symmetry. Similarly, we stress that if scattering processes can thermalize inflaton particles, they will follow a WIMP-like cosmological evolution.

We must finally take into account that $\phi$-particles may be produced by the oscillating inflaton field through the non-linear $\lambda\phi^4$ interaction. In each oscillation, $\phi$-particles become lighter than the field's oscillation frequency, $\omega_\phi\sim m_{\phi,\text{eff}}$, which kinematically allows their pair production at a rate $\Gamma_{\phi \rightarrow \delta \phi \delta \phi} = 0.17 \lambda^{3/2} \phi$ \cite{Kainulainen:2016vzv} (see also \cite{Ichikawa:2008ne, Cosme:2018nly}). This falls more slowly than expansion during the radiation era, since $\phi\propto H^{1/2}$, and may therefore lead to the condensate's evaporation before it reaches the cold dark matter regime at $\phi \lesssim \phi_{\text{DM}}$, where it becomes kinematically forbidden.

The produced inflaton particles are relativistic, since they have typical momenta $|\mathbf{p}|\sim \omega_\phi\sim \sqrt{\lambda}\phi$ and mass $m_\phi \ll \omega_\phi$ once the condensate evaporates. They are also decoupled from the radiation bath since they only interact with the latter through the heavy $\psi_{1,2}$ fermions, so they will simply redshift as dark radiation until they become non-relativistic, i.e.~for $\Delta N_e\sim \log (\omega_\phi/m_\phi)\sim \log(\phi/\phi_{\text{DM}})$. Accordingly, they behave as dark radiation for essentially the same amount of time as the oscillating inflaton condensate had it not evaporated. This implies that the cosmological evolution of the produced inflaton particles is indistinguishable from that of the oscillating background field, and for simplicity we will henceforth consider the latter picture. 

The ratio $n_\phi/s$ remains constant in the cold dark matter regime, for $\phi<\phi_{\text{DM}}$, being given by:
\begin{equation}
\begin{aligned}
\frac{n_\phi}{s} = \frac{\rho_{\phi}/m_{\phi}}{(2\pi^2/45)  g_*(T_{\text{DM}}) T_{\text{DM}}^3} = \sqrt{\frac{\lambda}{2}} \frac{45}{2 \pi^2} \frac{1}{g_{*}} \left(\frac{\phi}{T}\right)^3 \quad,
\end{aligned}
\end{equation}
where we used that $(\phi/T)g_*^{-1/3}$ remains constant throughout the dark radiation phase. The present dark matter abundance $\Omega_c \approx 0.25$ \cite{Aghanim:2018eyx} then yields the following estimates for the inflaton mass and the temperature at which it starts behaving as cold dark matter:
\begin{equation}
	\begin{aligned}
	m_{\phi} \approx 10^{-3} \left(\frac{\Omega_c}{0.25}\right) \left(\frac{g_*}{106.75}\right)\left(\frac{10^{-15}}{\lambda}\right)^{1/2}\left(\frac{10^4}{\phi/T}\right)^{3} \text{ eV} \:, \\
	T_{\text{DM}} \approx 11 \left(\frac{\Omega_c}{0.25}\right) \left(\frac{g_*}{106.75}\right)^{4/3}\left(\frac{10^{-15}}{\lambda}\right)\left(\frac{10^4}{\phi/T}\right)^{4}  \text{ eV} ~.
	\end{aligned}
\end{equation}
For the parameter space consistent with $50-60$ e-folds of inflation, the observed dark matter abundance is attained for $m_\phi\sim 10^{-4}-10^{-1}$ eV and $T_{\text{DM}}\sim 5-10^4$ eV, above the temperature of matter-radiation equality. The inflaton mass must thus be small so as to not overclose the Universe, but this is technically natural given the absence of radiative corrections to the scalar potential.

Given its light mass and feeble interactions, both direct and indirect searches seem unlikely to find inflaton dark matter in the near future. However, its inflationary origin and post-inflationary evolution can provide testable observational signatures. The first is the presence of cold dark matter isocurvature perturbations in the CMB spectrum, parametrized by
\begin{equation}
	S_c = -3 H \left( \frac{\delta \rho_c}{\dot{\rho}_c} - \frac{\delta \rho_r}{\dot{\rho}_r} \right) \quad.
\end{equation}
These are fully determined by the spectrum of small thermal inflaton fluctuations about the background $\phi$ generated during the warm inflationary phase. Hence, $S_c$ reads
\begin{equation}
S_c = \left( 2 - \frac{12 Q}{3+5Q} \right) \frac{\phi'}{\phi} \mathcal{R} \label{Isocurvature Perturbations}\quad,
\end{equation}
where $\mathcal{R} = \delta \phi/\phi'$ is the gauge-invariant (adiabatic) curvature perturbation and all quantities are evaluated when the relevant CMB scales become super-horizon during inflation. Since $\phi'<0$, isocurvature modes are anti-correlated with the main adiabatic component for $Q_*<3$, the regime where the WLI model with a quartic potential is technically and observationally consistent \cite{Bastero-Gil:2016, Bastero-Gil:2017wwl, Bastero-Gil:2018uep}.

Their contribution to the primordial perturbation spectrum can be parametrized by the ratio $\beta_{\text{Iso}} = B_c^2 /(B_c^2 + 1)$ with $B_c = S_c/\mathcal{R}$. For the example in Figs.~\ref{cosmological_evolution} and \ref{inflationary_dynamics}, $\beta_{\text{Iso}} = 3.15 \times 10^{-4}$, and in the consistent parameter range we find $\beta_{\text{Iso}} \in [3,4] \times 10^{-4}$, below the Planck bounds on fully anti-correlated isocurvature perturbations, $\beta_{\text{Iso}} < 2 \times 10^{-3}$ \cite{PlanckCI:2018}. Since $Q$ and $\phi$ are functions of the number of e-folds, the isocurvature spectral index, $n_I \equiv 1 + d \log \Delta_I^2/dN_e$, differs from the adiabatic one. For instance, in our working example we find $n_I \approx 0.999$ and $n_s \approx 0.965$. Future searches of cold dark matter isocurvature modes with the CMB-S4 mission \cite{CMB-S4:2016} will therefore be crucial to test our scenario.


The second is the prediction of extra relativistic degrees of freedom during nucleosynthesis (BBN), whose main contribution is given by the oscillating condensate,  $\Delta N_{\text{eff}} =  4.4 \rho_{\phi}/\rho_{\gamma}$ \cite{Baumann:2018}, with $\rho_{\gamma}$ denoting the photon energy density, which is fixed by the value of $\phi/T$ at the onset of inflaton oscillations and the thermal evolution of the SM DOF. For the case represented above, $\Delta N_{\text{eff}} \approx 0.13$. BBN bounds on $\Delta N_{\text{eff}}$, $\Delta N_{\text{eff}} <0.20$ \cite{Cyburt} are already strong enough to restrict the parameter space consistent with $\sim$50-60 e-folds of inflation to $g \sim 0.1 - 0.2$, $h \sim 1 - 2$, $M \sim 10^{14}-10^{15}$ GeV and $Q_* \sim 10^{-3} - 10^{-2}$.

This contribution vanishes before recombination, once the condensate $\phi$ starts behaving as cold dark matter. The small inflationary thermal fluctuations about $\phi$ remain, however, relativistic until after recombination, since $m_{\phi} \lesssim 0.1 \text{ eV}$. Although these consequently give only a subdominant contribution to the present cold dark matter abundance, they change the number of relativistic DOF at both BBN and recombination by $\Delta N_{\text{eff}}^{\text{th}} = (4/7) \left(43/4g_*\right)^{4/3} \approx 0.027$, where $g_*\approx 106.75$ at the end of inflation
\cite{Aghanim:2018eyx}, which may be probed with the next generation CMB experiments and large-scale structure surveys \cite{Baumann:2018}.

Finally, we note that warm inflation itself has distinctive observational features, namely a modified consistency relation between the tensor-to-scalar ratio and the tensor spectral index \cite{Cai:2010wt,Bartrum:2013fia} and non-Gaussianity with a characteristic bispectrum \cite{Bastero-Gil:2014raa}. All these observables thus provide, in conjunction, a very distinctive and testable scenario for inflaton-dark matter unification, where one can probe the inflationary nature of dark matter even if it interacts too feebly with known particles to be found in direct searches or collider experiments.

The scenario described in this Letter shows that inflaton-dark matter unification is a natural feature in warm inflation. The underlying symmetries of the model allow the inflaton to decay significantly only during inflation, remaining essentially stable throughout the remainder of the cosmic history. The basic ingredients of the model can nevertheless accommodate more general scenarios, with different forms of the scalar potential and further interactions that could allow for further (partial) inflaton decay after inflation or even its thermalization with the cosmic heat bath. Our model may also, in principle, be embedded within different SM extensions, which could potentially yield novel observational signatures of inflaton-dark matter unification.


\vspace{0.1cm}
\begin{acknowledgments}
L.B.V is supported by FCT grant PD/BD/140917/2019. J.G.R. is supported by the FCT Investigator Grant No. IF/01597/2015. This work was partially supported by the H2020-MSCA-RISE-2015 Grant No. StronGrHEP-690904 and by the CIDMA Project
No. UID/MAT/04106/2019.
 \vfill
\end{acknowledgments}



\begin{thebibliography}{99}
\bibitem{note3a} Our results are derived under the technical assumptions of adiabaticity, near thermal equilibrium and flat space (at scales $\sim T^{-1}$), which restrict the parameter space of the model, although without any fine-tuning. We note, however, that outside this parametric regime, non-equilibrium dissipative effects will still be present and need not be suppressed, so that one is not necessarily in a cold inflation regime.
\bibitem{note4a} This depends, of course, on how the WLI model is embedded within an extension of the Standard Model.
\end{thebibliography}
\end{document}